\documentclass[aps,pre,showpacs,superscriptaddresss]{revtex4}
\usepackage[]{graphicx}
\usepackage{bm}

\begin{document}

\title{Probabilistic Quantum Control Via Indirect Measurement}
\author{A.~Mandilara and J.~W.~Clark}
\affiliation{Department of Physics, Washington University, 
Saint Louis, MO 63130}

\date{\today}

\begin{abstract}
The most basic scenario of quantum control involves the 
organized manipulation of pure dynamical states of the system 
by means of unitary transformations.  Recently, Vilela Mendes
and Mank'o have shown that the conditions for controllability
on the state space become less restrictive if unitary control
operations may be supplemented by projective measurement.  The 
present work builds on this idea, introducing the additional element 
of indirect measurement to achieve a kind of remote control.  
The target system that is to be remotely controlled is first 
entangled with another identical system, called the control 
system. The control system is then subjected to unitary 
transformations plus projective measurement.  As anticipated
by Schr\"odinger, such control via entanglement is necessarily
probabilistic in nature.  On the other hand, under appropriate 
conditions the remote-control scenario offers the 
special advantages of robustness against decoherence and a greater 
repertoire of unitary transformations.  Simulations carried out 
for a two-level system demonstrate that, with optimization of 
control parameters, a substantial gain in the population of 
reachable states can be realized.  
\end{abstract}
\maketitle

\section{Introduction}

The conditions under which a quantum-mechanical system
is controllable and the degree to which control is possible 
are issues of considerable theoretical and practical importance.
Many different definitions of controllability are currently
in play.  Let us suppose the time-development of the system is
described by a Schr\"odinger equation (with $\hbar = 1$)
\begin{equation}
i\frac{d}{dt}|\psi(t)\rangle
   = \left[H_{o}+\sum_{n=1}^{r} f_{n}(t)H_{n}\right]|\psi(t)\rangle \,, 
\end{equation}
where the $f_{n}(t)$ are independent, bounded, measurable 
control functions.  The most common notion of controllability is 
{\it pure-state controllability} \cite{schi}, taken to mean that
starting in any given pure state $|\psi_0=|\psi(t_0)\rangle$, there
exists a set of control functions $f_n(t)$ such that any pure 
final state $|\psi_f\rangle = |\psi(t_f) \rangle$ can be reached at 
some later time $t_f > t_i$.  This is equivalent to saying that 
there exists a set of control functions $f_n(t)$, a time 
$t_f > t_0$, and a unitary operator $U(t)$ satisfying 
\begin{equation}
i {d \over d t} U(t) = 
    \left[H_{o}+\sum_{n=1}^{r} f_{n}(t)H_{n}\right]U(t) \,,
\end{equation}
such that $U(t_0) |\psi_0 \rangle = |\psi_0 \rangle $ and
$U(t_f) | \psi_0 \rangle = | \psi_f \rangle $.  A stronger 
condition is {\it complete controllability}, in the sense that
any unitary operator $U$ is dynamically accessible from the 
identity operator.

The most incisive results are available for the restricted, but 
practically important, case of a system with a finite number 
of energy levels, more precisely, a system whose eigenstates span 
a Hilbert space with finite dimension $N$.  In particular, a 
necessary and sufficient condition for pure-state controllability 
\cite{schi} is that the dynamical Lie group $G({\cal A})$ generated 
by the set of operators $\{iH_0, iH_1, \ldots  iH_r  \}$ is 
equal to $U(N)$, $SU(N)$, or (if $N$ is even) either $Sp (N/2)$ 
or $Sp (N /2)\times U(1)$.  It may be shown that these conditions 
can only be satisfied if the dynamical Lie algebra of the system is 
$u(N)$, $su(N)$, or (if $N$ is even) either $sp (N/2)$ or 
$sp (N/2)\oplus U(1)$.  {\it Complete controllability} of the 
$N$-level problem is naturally more demanding: It is necessary 
and sufficient that $G({\cal A})$ coincide with the largest of 
the groups listed, namely $U(N)$.

We note that fundamental theorems on controllability were established
for a more general class of quantum systems at the very beginning
of the subject of quantum control \cite{huang,clark02,clark03}.  This 
class includes continuous systems with unbounded observables (e.g., position,
momentum, kinetic energy), whose states span an infinite-dimensional
Hilbert space.  The domain problems were dealt with by assuming
the existence of an analytical domain in the sense of Nelson \cite{nelson},
and available geometric methods for finite-dimensional bilinear control 
systems \cite{brockett,sussmann1,sussmann2,kunita} were adapted to derive 
controllability results in terms of certain Lie algebras.  In fact, 
theorems commonly stated for finite-level systems may be extracted 
as special cases of the results of Ref.~\cite{huang}.

The objective of this paper is expand the scope of control beyond
the implementation of unitary operators, exploiting the phenomenon
of entanglement and the option to carry out measurements on the
given  system (or its surrogate).  In the interest of 
transparency, we shall avoid troublesome domain problems by 
focusing on a quantum system described in a state space of 
finite dimension $N$.

We take as a starting point the recent result of Vilela Mendes and 
Man'ko \cite{mendes} establishing that in some situations,
a nonunitarily controllable system can be controlled by the joint
action of projective measurement plus unitary evolution.  More 
precisely:

\noindent
{\bf Theorem:} {\it For a specified target state $| \psi_f \rangle$,
there exists a family of observables $M[|\psi_f \rangle ]$ such that 
measurement of any one of them on an arbitrary initial state 
$|\psi_0 \rangle$, followed by unitary evolution, leads to 
$|\psi_f \rangle$ if $G({\cal A})$ is either $O(N)$ or $Sp(N/2)$.}

\noindent
As pointed out in Ref.~\cite{mendes}, the system is already
pure-state controllable if $G({\cal A}) = Sp(N/2)$, but it still
might be more efficient to use the measurement/evolution strategy.
Also, if both the initial and final states are fixed, pure-state
controllability may be achieved with this strategy even if
$G({\cal A})$ is a much smaller subgroups of $U(N)$ than 
$O(N)$ or $Sp(N/2)$.

Thus, the conditions required for controllability are weakened if 
unitary control is supplemented by projective measurement.  
However, when the measurement is performed on a given observable 
of the system, the possible outcomes are necessarily restricted 
to the set of eigenstates of this observable.  

The present work aims to overcome this limitation by extending the
hybrid measurement/unitary approach to control a step further, 
exploring the additional prospects opened by performing the 
measurement on an {\it entangled partner} of the system 
in question (cf. Ref.~\cite{lloyd}).  The basic scheme is introduced in Sec.~II.  As anticipated 
by Schr\"odinger \cite{schr} in 1935, a salient feature of this 
exploitation of entanglement is that the ``remote control'' so 
attempted can no longer absolute, but is instead probabilistic in 
character.  Nevertheless, an enlargement of the reachable set of 
states can be achieved.  Alternative algebraic and geometric 
descriptions of the proposed control scheme are presented in 
Sec.~III.  In Sec.~IV, we illustrate the possibilities opened by 
the remote-control strategy for the simple case of a two-level 
system ($N=2$) as realized, for example, by a Pauli spin~1/2.  
The efficacy of the method, measured by the number of reachable 
final states and the probability of a successful outcome, is 
tested in a simulation in which adjustable control parameters are 
optimized to minimize the distance of the actual state from the 
desired final state.  In Sec.~V we consider the effects of decoherence 
within the remote-control scenario.  As usual, the directly controlled 
system suffers from decoherence due to its environment, whereas the 
remotely controlled target system, kept isolated from its 
surroundings, remains immune.  We conclude in Sec.~VI with some
remarks on the genesis of the idea proposed here, and on its 
further development.

\section{Control via Entanglement}

The proposed control scheme -- control via indirect projective 
measurement -- involves three basic steps.  Two systems are 
involved: (i) the {\it target} $N$-level system, which we wish 
to move by means of indirect influences into a pre-selected 
final state and (ii) the {\it control} system, an identical, 
entangled partner of the target system which is directly 
steered or shoved by control operations from the available 
repertoire.  It is supposed that the target system is initially
in a pure state
\begin{equation}
|\psi^{(\rm t)} \rangle =  \sum_{i=1}^N c_i|e_i^{(\rm t)}\rangle \,, 
\end{equation}
expressed in a convenient basis $ \{| e_i^{(\rm t)} \rangle \}$. 
Likewise, the control system is initially in a pure state 
$| \psi^{(\rm c)} \rangle $ similarly expressed in its own
state space. 

{\it First}, we entangle the target system with the control system,
e.g.\ by means of a non-local two qubit operation.
The combined system undergoes the change
\begin{equation}
  |{\psi}^{(\rm t)} \rangle \otimes |{\psi}^{(\rm c)} \rangle 
  =  \sum_{i=1}^N c_i|e_i^{(\rm t)} \rangle 
     \otimes \sum_{l=1}^N b_l|e_l^{(\rm c)} \rangle\
     \stackrel{\cal E}{\longrightarrow}|\chi\rangle 
  = \sum_{i=1}^N a_i|e_i^{(\rm t)}\rangle \, |e_i^{(\rm c)} \rangle  \,,
\end{equation}
where ${\cal E}$ symbolizes entanglement and we suppress the tensor 
product notation in the third member.
In the density-matrix formulation, the partial density matrix of
the target system, obtained by tracing over the control system,
undergoes the transformation
\begin{equation}
(\rho^{(\rm t)})_{ij}= a_i a^*_j |e_i^{(\rm t)}\rangle
\langle e_j^{(\rm t)}
   |\stackrel{\cal E}{ \longrightarrow} (\rho^{(\rm t)})_{ij}
  =|a_i|^2 |e_i^{(\rm t)} \rangle\langle e_j^{(\rm t)}| \delta_{ij} \,.
\end{equation}

{\it Second}, one of the available unitary transformations $U(t)$ 
is applied to the control system thus:
\begin{equation}
  |\chi' \rangle 
 = \sum_{i,j=1}^N U_{ji} a_i|e_i^{(\rm t)}\rangle\, |e_j^{(\rm c)} \rangle\,.
\end{equation}
While $\rho^{(\rm t)}$ remains unaffected, the partial density 
matrix $\rho^{(\rm c)}$ of the 
control system begins a forced evolution according to
\begin{equation}
(\rho^{(\rm c)})_{ij} =|a_i|^2|e_i^{(\rm c)}\rangle
\langle e_j^{(\rm c)}| \delta_{ij}\rightarrow ({\rho^{(\rm c)}}')_{ij} = 
\sum_{k=1}^N U_{ik}|a_k|^2 U_{kj}^* |e_i^{(\rm c)} \rangle 
\langle e_j^{(\rm c)}|\,.
\end{equation}

In the {\it third} and final step, a projective measurement is performed on 
the control system for a selected observable $X$.  Without loss of
generality, we may assume that the basis $\{ |e_i^{(\rm c)}\rangle\}$ 
in the state space of the control system is an eigenbasis of the 
chosen observable, which may then be expressed as 
\begin{equation}
X = \sum_{n=1}^{N}x_{n}| e_n^{(\rm c)} \rangle \langle e_n^{(\rm c)} | \,. 
\end{equation}
The measurement will then yield the eigenvalue $x_m$ of $X$ with
probability 
\begin{equation}
P_m=\sum_{i=1}^NU_{mi}|a_i|^2U_{im}^*\,, 
\end{equation}
leaving the combined system in a state that is no longer entangled, namely
\begin{equation} 
|\chi_m'' \rangle 
 = \frac{1}{\sqrt{P_{m}}} \sum_{k=1}^N U_{mk}a_k 
|e_k^{(\rm t)} \rangle \otimes |e_m^{(\rm c)}\, \rangle \,.
\end{equation} 

It is seen that the final state of the target system is in general 
a {\it superposition} of eigenstates of the observable $X$, rather
than the particular eigenstate corresponding to the result of measurement,
as it would be in a simple {\it direct} projective measurement.
Furthermore, there are $N(>1)$ possible results of the three-step 
control procedure, which therefore assumes a probabilistic character.
As we shall see, the advantage of certainty of outcome is traded for
a potentially expanded range of control.
Another positive aspect of remote control is that it can
overcome the limitation of unitary control to transformations
of the state of the target system within a restricted equivalence
class determined by the set of eigenvalues of the initial
density matrix \cite{schi}.
\section{Algebraic and Geometric Descriptions of Indirect Control}

\subsection{Algebraic Treatment}

The total effect of the indirect control scheme on the target system
can be represented in terms of a set of $N$ diagonal matrices
$\Upsilon_m = \left( U_{ml} \delta_{ln}  \right)$ representing Kraus
operators, one for each of the $N$ possible results of the measurement 
performed on the control system.  Due the unpredictability of the 
final state, the property of controllability, as strictly defined, 
does not apply to the target system.

This situation contrasts with what is found in the theory of 
universal quantum interfaces developed in Ref.~\cite{landahl}, where 
similar schemes involving remote control are formalized, but with 
broader intent within the contexts of quantum computation and 
quantum communication.  In that work, the target system is shown 
to be both controllable and observable through control and 
observation of the control bit to which it is coupled. The main 
distinction between the two approaches is the following. In 
Ref.~\cite{landahl}, the control and target systems remain in
close proximity and the interaction between them can have indefinite
duration, whereas in the remote-control scenario envisioned here,
the systems are in transient interaction, and then separate from one
another.  In some circumstances, the disjunction of the two systems
may prove desirable or advantageous.

Controllability being moot, our consideration turns to reachable
sets of the target system.  It is easily seen that if the control
system is controllable, then every state of the target system is
reachable.  Every unitary transformation $U$ of the target system
is available for temporal manipulation of the control system.  To 
each of these there correspond $N$ non-unitary transformations 
$\Upsilon_m$, and the mapping between $U$ and each of the $\Upsilon_m$ is 
one-to-one.  Hence, every state of the target system is reachable.

If the control system is not controllable, then every state of
the target system may or may not be reachable.  To illustrate this
possibility, consider the case in which both the control and target
systems are spin-$1/2$ particles.  Let the unitary transformations
available for application be specified by
\begin{equation}
U(\theta, \phi)=\left( \begin{array}{ccc}
\cos(\theta/2) & -\sin(\theta/2)e^{i\phi} \\
\sin(\theta/2) & \cos(\theta/2)e^{i\phi}
\end{array} \right)\,,
\end{equation}
where $ 0\leq\phi < \pi$ and $ 0\leq\theta\leq \pi/2$.  In this
case, the set of the states reachable from an initial state
on the equator of the Bloch sphere covers only two quadrants 
of the Bloch sphere.

Let us apply these transformation to a control system that
is maximally entangled with the target system, and then perform
a measurement on the spin component $\sigma_z$  of the
control system.  The available transformations are then expanded to
\begin{equation}
\Upsilon_1(\theta, \phi)=\left( \begin{array}{cc}
\cos(\theta/2) & 0 \\ 0 & -\sin(\theta/2)e^{i\phi} \end{array} \right)\,,
\qquad \Upsilon_2(\theta, \phi)=\left( \begin{array}{cc}
\sin(\theta/2) & 0 \\0 & \cos(\theta/2)e^{i\phi} \end{array} \right)\,.
\end{equation}
Thus, the reachable set for the target system is the whole Bloch sphere,
even if the set reachable by applying only the specified $U$ transformations
is just two quadrants.  We note that the assumed condition of maximal 
entanglement simplifies the proof but is not essential.  This possibility
for enlargement of the reachable set is illustrated in the optimization
problem solved in the next section.
                  
Sequential application of the probabilistic remote-control scheme
is not in general effective in further extension of the range of 
control. The $\Upsilon_m$ matrices are diagonal and
necessarily commute with one another; consequently, the 
advantages of a Lie algebra do not apply.  Unlike unitary 
operators, Kraus operators are not guaranteed the property
that they can be combined to give new directions of control
in the state space \cite{viola}.  

Finally, if $\Upsilon_m$ operations are combined with unitary operations 
on the target system, the commutativity is lifted and the repertoire of
available transformations on the target system is enlarged.
Again we chose a two-level example to illustrate the point.
Suppose the only available quantum gate for the system is 
the Hadamard gate
\begin{equation}
U_H = \frac{1}{\sqrt{2}}\left[ \begin{array}{cc}
                     1 & 1 \\
                     1 & -1 \end{array} \right] \,.
\label{had}
\end{equation}
Then upon implementing the probabilistic remote-control scheme for
this target system (involving entanglement, application of the
 Hadamard gate on the control qubit, and projective measurement), 
we obtain an additional gate
\begin{equation}
\Upsilon_2 = Z = \left[ \begin{array}{cc}
                     1 & 0 \\
                     0 & -1 \end{array} \right] \,.
\end{equation}
By successive applications of $U_H$ and $Z$ we further extend
the set of reachable states.  In this case it happens $\Upsilon_1=I$ 
and $\Upsilon_2 = Z $ are both unitary, but it will not generally
be the case that all the $\Upsilon_m$ are unitary.  It is 
interesting to note that the probabilistic character of the
control scheme can be overcome by applying unitary transformations
on the target system so as to feed back the indirect measurement
results, as proposed in Ref.~\cite{viola}.

\subsection{Geometric Treatment}

Again for the sake of simplicity and clarity, we consider a two-level
system ($N=2$).  Physically, the system might be a single Pauli
spin-1/2 or a two-level atom, having energy eigenstates
denoted $ | 0 \rangle $ and $ | 1 \rangle $.
  
The geometric description is based on the coherent-vector (or Bloch-vector) 
picture of quantum dynamics
\cite{alicki,emma}.
 The coherent vector $\bf v$ can  
represent a pure state on the Bloch sphere as well as a mixed state 
lying in the interior of the sphere.  Its magnitude, or length, is 
defined by 
$ \| {\bf v}\| = ({\rm tr}{\rho}^{2}-1/2)^{1/2}$, 
and its Cartesian components by 
$v_{x}=2^{-1/2}{\rm tr}(\rho\sigma_{x})$, 
$v_{y}=2^{-1/2}{\rm tr}(\rho \sigma_{y})$, and 
$v_{z}=2^{-1/2}{\rm tr}(\rho \sigma_{z})$.
Whether it refers to an entangled or non-entangled quantum system,
the coherent vector $\bf v$ evolves with time according to the
following rules.  
\quad
\begin{itemize}
\item[(a)]
The coherent vector may shrink in magnitude, i.e., contract to a 
shell of smaller radius, if and only if the system becomes entangled. 
The tip of the vector traces a continuous path, namely a line passing 
through the initial position and perpendicular to the axis that connects 
the einselected states~\cite{pente} [see Fig.~1(a)].  Either
premeasurement~\cite{pente} or decoherence will drive the coherent 
vector in this manner, because both these processes imply entanglement.
\item[(b)]   
A unitary transformation leaves ${\rm tr}{\rho}^2$ invariant and hence 
does not change the magnitude of the coherent vector $\bf v$. 
Accordingly, a unitary transformation can only rotate $\bf v$
on the shell of radius equal to $\| {\bf v}\| $ 
[see Fig.~1(b)]. The effect of the rotation is independent of 
the magnitude of the vector.
\item[(c)]  
A mixed state may become pure if the entangled state becomes 
disentangled and the bipartite system becomes separable. 
This can occur through projective measurement on one of the 
two systems.  Both of the systems are purified, but not in a 
deterministic manner. The possible final states depend on the 
observable that is measured and on the entangled state.
Figure~1(c) gives a simple example in which the projective
measurement is made on an observable whose eigenstates coincide 
with the Schmidt basis of the measured system.
\end{itemize}
\begin{figure}
\vspace*{8.0cm}
\includegraphics{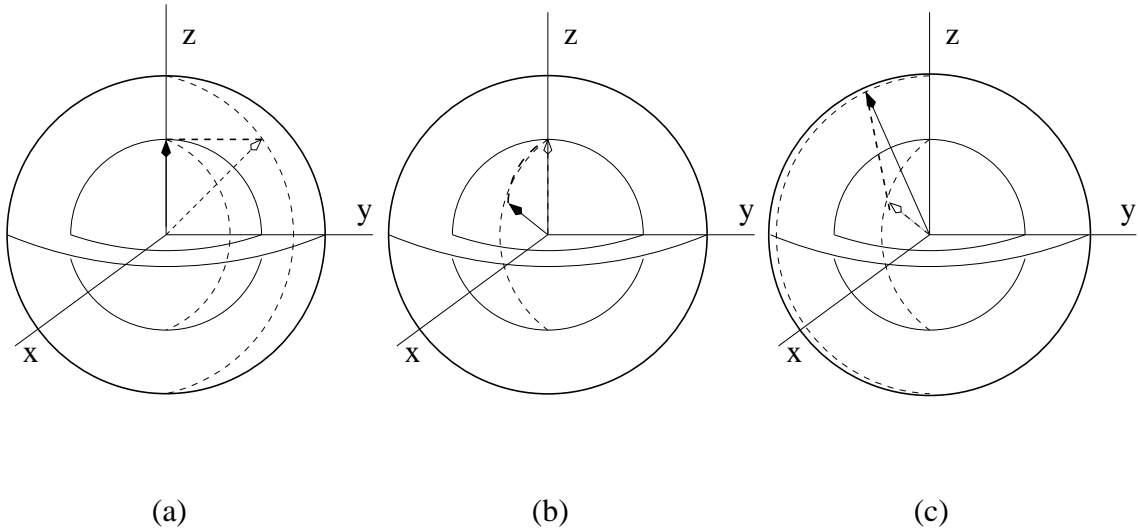}
\vspace{5.8cm}
\caption{Coherent-vector representation of quantum dynamics for
a two-level system.
(a) The effect of entanglement on a pure state.  (b) Unitary
transformation of a mixed state.  (c) Projective measurement 
on either system of an entangled pair.  In each operation, the
initial vector [final vector] is drawn as an arrow with a 
white [black] head.}
\end{figure}

Probabilistic, indirect quantum control, as introduced in Sec.~II, 
involves all three of these operations.  The geometric description of this
process is illustrated in Fig.~2 for the $N=2$ case (which is
actually the only nontrivial case that one can draw.)   Reiterating,
the scheme is to:
\begin{itemize}
\item[(a)]
Entangle the target system with the control system.  This causes
shrinkage of the coherent vectors of both systems [Fig.~2(a)].
\item[(b)]
Apply a unitary transformation to the control system. The 
coherent vector ${\bf v}^{({\rm c})}$ of the control system
rotates without change of magnitude, while the coherent
vector ${\bf v}^{({\rm t})}$ of the target system is 
unaffected by the transformation [Fig.~2(b)]. 
\item[(c)]
Make a projective measurement on the control system.  The final 
coherent vectors are not determined.  One has either 
${\bf v}_1^{({\rm c})}$ or ${\bf v}_2^{({\rm c})}$ for the control
system, ${\bf v}_1^{({\rm t})}$ or ${\bf v}_2^{({\rm t})}$ for 
the target [Fig.~2(c)].  The initial and final coherent vectors 
obey a set of angle rules; in particular 
\begin{eqnarray}
\angle \left({\bf v}^{({\rm c})},{\bf v}^{({\rm t})} \right) 
&=& \angle \left({\bf v}^{({\rm c})},{\bf v}_1^{({\rm c})} \right)\,,
\angle \left({\bf v}^{({\rm c})},{\bf v}^{({\rm t})} \right) 
= \pi - \angle \left({\bf v}^{({\rm c})},{\bf v}_2^{({\rm c})} \right)\,,
\nonumber \\
\angle \left({\bf v}^{({\rm c})}_1,{\bf v}^{({\rm t})}_1 \right) 
&=& \angle \left({\bf v}^{({\rm t})},{\bf v}_1^{({\rm t})} \right)\,,
\qquad  
\angle \left({\bf v}^{({\rm c})}_2 ,{\bf v}^{({\rm t})}_2 \right) 
= \pi - \angle \left({\bf v}^{({\rm t})},{\bf v}_2^{({\rm t})} \right)\,.
\end{eqnarray}
Also, $\angle ( {\bf v}^{({\rm c})}_1, {\bf v}^{({\rm c})}_2)= \pi$, while 
$\angle ( {\bf v}^{({\rm t})}_1, {\bf v}^{({\rm t})}_2)$ depends 
on the initial state of the target system and the unitary transformation 
applied to the control system.
In the special case where the two systems are maximally entangled
then $\angle ( {\bf v}^{({\rm t})}_1, {\bf v}^{({\rm t})}_2)= \pi$.
\end{itemize}
\begin{figure}
\vspace*{8.0cm}
\includegraphics{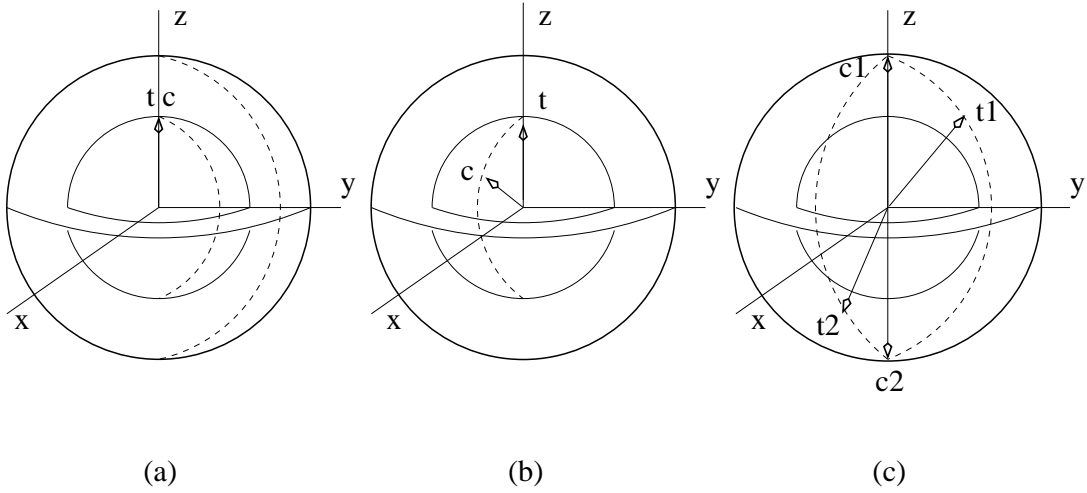}
\vspace{5.8cm}

\caption{(a) Control and target system become entangled. 
         (b) A unitary transformation is applied to the control system.
         (c) A projective measurement is performed on the control system.}
\end{figure}

With exclusive use of unitary transformations to control a system,
the coherent vector is rigorously confined to the shell of the
Bloch sphere.  Probabilistic remote quantum control
permits the coherent vector to move to the interior of the 
sphere as well, thereby opening new pathways to the desired 
final state.

\section{Optimized probabilistic control: A simulation}
The benefits (and drawbacks) of the probabilistic remote-control 
process are exemplified in a problem drawn from nuclear magnetic
resonance.  If there is a constant magnetic field of strength
$B_0$  present along the $z$-axis, the Hamiltonian of a 
spin-1/2 particle is $H_{o}=\omega \sigma_z = \omega Z$, and the 
time evolution operator for the system is given by
\begin{equation}
 U_0 (t) = \left( \begin{array}{cc}
           e^{-i\omega t} &0 \\
           0    &   e^{i\omega t}
           \end{array} \right)\,.
\end{equation}

If a resonant magnetic field $(B_x,B_y)$ is also applied in
the $x-y$ plane, we have an additional time-dependent gate
\begin{equation}
 U_1 (t) = \left( \begin{array}{cc}
           \cos(g \ t) &-i\ \sin(g\ t) \\
           -i\ \sin(g\ t)  &  \cos(g \ t)
           \end{array} \right)\,.
\end{equation}
Suppose we wish to reach a particular final state at the exact 
time $T$, by applying $U_0 (T/2)$ and then $U_1 (T/2)$.  The 
parameters available for adjustment (``optimization'') are 
the field strengths $B_0$, $B_x$, and $B_y$, or more precisely 
$\omega$ and the coupling constant $g$. Here we note the precedent set by Ref. \cite{emma}
 in organizing pure-state control of a two-level quantum system within the
 geometric intepretation on the Bloch sphere.

Simulations were performed to test the efficacy
of two different control schemes, namely unitary control
alone and the probabilistic remote-control scenario.
A hundred random pairs of initial states were chosen
and simulations were performed for both control schemes.
 With the initial time at 0, the numerical experiment
was repeated for ten different final times $T$,
keeping the adjustable parameters within the ranges
  $0\leq g\leq 2\pi$ and $0\leq\omega\leq 2\pi$.  
The two relevant performance measures are the fraction of
final states successfully reached and the overall probability
of reaching the final state of a pair.  The results of
averaging over all simulations are shown in Table I.  As might 
be expected, the fraction of target states successfully reached 
is significantly larger , more than double when the indirect-measurement 
protocol is implemented.  However, this advantage is compensated 
by the probabilistic nature of the remote-control process, 
such that the overall success rates for the two methods are similar.
                                                                                                                             
\vskip .6truecm
                                                                                                                             
\noindent
{\bf Table I.} Comparison of probabilistic remote control with pure
unitary control, for a spin-1/2 system
\vskip .4truecm
\parindent 70pt
                                                                                                                             
\begin{tabular}{|c|c|c|c|}
\hline\hline
~Control ~&~  Number of  ~&~  Target final ~&~ Net Probability ~\\
~protocol  ~&~   pairs tested ~&~  states reached ~&~ of success ~\\
\hline
~Unitary ~&~ 100 ~&~ $2.4\pm0.4$ ~&~$ 0.024\pm0.004$~ \\
~Remote  ~&~ 100 ~&~ $6.6\pm0.5$ ~&~$ 0.0345\pm0.0012$~ \\
\hline\hline
\end{tabular}
\parindent 20pt

\section{Role of decoherence}

Suppose that we entangle a pair of identical (sub)systems such 
that combined system is described by the state vector $| \psi \rangle $.
Now, arrange that the two subsystems become separated, such that
the target system, which is to be remote-controlled, is kept 
isolated from the environment, while the control system remains 
exposed in the laboratory, where we can perform unitary operations 
or measurements upon it.  The  control system soon interacts 
with the laboratory environment and becomes entangled with it;
schematically,
\begin{equation}
   |\psi\rangle
 = \sum_{i=1}^{N}a_{i}|e_{i}^{({\rm t})}\rangle \,
   |e_{i}^{({\rm c})} \rangle \otimes | {\rm env} \rangle 
   \stackrel{{\cal E}}{ \rightarrow} |\psi\rangle
 = \sum_{i=1}^{N}a_{i}|e_{i}^{({\rm t})}\rangle \,
   |e_{i}^{({\rm c})}\rangle \,|{\bf\epsilon}_{i}\rangle \,,
\end{equation}
where ${|{\bf\epsilon}_{i}\rangle}$ is a basis for the environment.

The entanglement between the target and control system is not affected 
by the presence of the environment \cite{yu}, and the statistical 
properties of the target system remain the same, i.e.,
\begin{equation}
(\rho^{({\rm t})})_{ij}
= \sum_{a=1}^{N}|a_{i}|^{2}|e_{i}^{({\rm t})}\rangle\langle 
e_{i}^{({\rm t})}|\delta_{ij}\,.
\end{equation}
Following the remote-control scenario, we next apply a unitary 
transformation on the control system, to obtain
\begin{equation}
  |\psi' \rangle
= \sum_{i,j=1}^{N}U_{ji}a_{i}|e_{i}^{({\rm t})}\rangle \,
  |e_{j}^{({\rm c})}\rangle\,|{\bf \epsilon}_{i}\rangle \,.
\end{equation}
However, the environment is still present and becomes entangled with
the new state of the control system:
\begin{equation}
|\psi'' \rangle
 = \sum_{i,j=1}^{N}U_{ji}a_{i}|e_{i}^{({\rm t})}\rangle \,
   |e_{j}^{({\rm c})}\rangle\,|{\bf \epsilon}_{j}\rangle \,.
\end{equation}
Finally, a projective measurement is performed on the control system, 
yielding
\begin{equation}
|\psi''' \rangle
=\frac{1}{\sqrt{P_{m}}}\sum_{i=1}^{N}a_{i}U_{im} 
|e_{i}^{({\rm t})}
\rangle\, |e_{m}^{({\rm c})}\rangle\,|{\bf \epsilon}_{m}\rangle 
=\frac{1}{\sqrt{P_{m}}}\sum_{i=1}^{N}a_{i}U_{im}|e_{i}^{({\rm t})}\rangle
  \otimes |e_{m}^{({\rm c})}\rangle\, |{\bf\epsilon}_{m}\rangle \,.
\end{equation}
We observe that the same results are obtained for the target system
as in the case where the environment is absent, whereas the control
system feels the effects of decoherence.

\section{Summary and prospects:  Remote control on entangled pairs}

Taking inspiration from quantum teleportation \cite{bennett} and 
from prior work of Vilela-Mendes and Man'ko \cite{mendes} in which
unitary control is supplemented by projective measurement, we 
have introduced a strategy for indirect control (``remote control'') 
of a target system through projective measurement on its entangled
partner.  We have thereby contributed to an ongoing unification 
of concepts and mathematical techniques developed in the fields 
of quantum control \cite{clark02,clark03} and quantum information 
theory \cite{nielsen}. The integration of these two thrusts began 
in 1995 with Lloyd's demonstration \cite{lloyd2} that ``almost any 
quantum logic gate is universal'' -- shorthand for the fact that 
universality in quantum computation can be achieved by repeated 
application of almost any two-level gate and a single-qubit gate.  
The proof of this statement rests on Lie-algebraic arguments that 
have long been a staple of geometric control theory.

Reversing the flow of ideas, we have exploited entanglement 
together with the option of projective measurement to enlarge
the scope of quantum control beyond what is attainable with
unitary transformations on system states.  Under the 
remote-control protocol, some states that were unreachable via 
simple unitary control now become reachable.  However, this advantage 
is tempered by the fact that the outcome of the final measurement 
operation is necessarily probabilistic, i.e., the outcome of 
remote control is described by a probability distribution over 
a set of quantum states.

Our attention here has been focused on the advantages that probabilistic
control via indirect measurement may offer in the manipulation of
a system occupying a single, initially pure quantum state.  As is 
evident, the idea may be extended to initial states of subsystems 
of a larger system, which in general are not pure and must be 
represented as density matrices.

Let the target system be an entangled bipartite system 
$({\rm t}_a, {\rm t}_b)$ described by 
\begin{equation}
 |\psi^{(t)}\rangle
 = \sum_{i=1}^{N}a_{i}|e_{i}^{({\rm t}_a)}\rangle 
  |e_{i}^{({\rm t}_b)}\rangle\,.
\end{equation}
The degree of entanglement of the system in this state may be 
quantified in terms of von Neumann entropy of the subsystem, or  more simply
the Schmidt number \cite{nielsen}.  Whatever appropriate measure
is chosen, it cannot be changed by applying a unitary transformation 
on either of the two subsystems.
However, the same is not true for the transformation accomplished
by remote control, which, for example, is capable of changing the 
Schmidt number of the bipartite system as we go from 
\begin{equation}
  |\chi\rangle
 = \sum_{i=1}^{N}a_{i}|e_{i}^{({\rm t}_a)}\rangle 
   |e_{i}^{({\rm t}_b)}\rangle |e_{i}^{({\rm c})}\rangle
\end{equation}
to
\begin{equation}
 |\chi'\rangle
 = \sum_{i,j=1}^{N}a_{i}U_{ji}|e_{i}^{({\rm t}_a)}\rangle 
   |e_{i}^{({\rm t}_b)}\rangle |e_{j}^{({\rm c})}\rangle
\end{equation}
to
\begin{eqnarray}
|\chi'' \rangle &=&\frac{1}{\sqrt{P_m}}\sum_{i=1}^{N}a_iU_{mi} 
|e_i^{({\rm t}_a)} \rangle |e_i^{({\rm t}_b)}\rangle 
|e_m^{({\rm c})}\rangle \nonumber \\
&=&\frac{1}{\sqrt{P_m}}\sum_{i} a_i U_{mi}|e_i^{({\rm t}_a)}\rangle
|e_i^{({\rm t}_b)}\rangle \otimes |e_m^{({\rm c})} \rangle \,.
\end{eqnarray}
In future work, the scheme proposed here will be applied to systems
that are entangled with many degrees of freedom.  In pursuing
such an investigation, one would like to determine the extent
to which non-unitary control operations can be used to counteract
undesirable effects arising from interactions between the system
and its environment.

\normalsize
\section*{Acknowledgments}
This research was supported by the U.~S.\ National Science Foundation 
under Grant No.~PHY-0140316 and by the Nipher Fund. J.W.C. would also
like to acknowledge partial support from FCT POCTI, FEDER in Portugal
and the hospitality of the Centro de Ci\^{e}ncias Mathem\'{a}ticas at
the Madeira Math Encounters.

\end{document}